# EXPRESSION OF MHC II GENES

Gorazd Drozina[1], Jiri Kohoutek[1], Nabila Jabran-Ferrat[2,*] and B. Matija Peterlin[1,*]

[1]*Departments of Medicine, Microbiology and Immunology*

*Rosalind Russell Medical Research Center*

*University of California, San Francisco*

*San Francisco, CA 94143*

[2]*Structural Immuno-Pharmacology*

*Institute of Pharmacology and Structural Biology*

*CNRS VMR 5089*

*31400 Toulouse*

*France*

[*]Correspondence: matija@itsa.ucsf.edu
Nabila.Jabrane-Ferrat@ipbs.fr

ABSTRACT

Innate and adaptive immunity are connected via antigen processing and presentation (APP), which results in the presentation of antigenic peptides to T cells in the complex with the major histocompatibility (MHC) determinants. MHC class II (MHC II) determinants present antigens to CD 4+ T cells, which are the main regulators of the immune response. Their genes are transcribed from compact promoters that form first the MHC II enhanceosome, which contains DNA-bound activators and then the MHC II transcriptosome with the addition of the class II transactivator (CIITA). CIITA is the master regulator of MHC II transcription. It is expressed constitutively in dendritic cells (DC) and mature B cells and is inducible in most other cell types. Three isoforms of CIITA exist, depending on cell type and inducing signals. CIITA is regulated at the levels of transcription and post-translational modifications, which are still not very clear. Inappropriate immune responses are found in several diseases, which include cancer and autoimmunity. Since CIITA regulates the expression of MHC II genes, it is involved directly in the regulation of the immune response. The knowledge of CIITA will facilitate the manipulation of the immune response and might contribute to the treatment of these diseases.

# INNATE AND ADAPTIVE IMMUNITY

The immune system is composed of innate and adaptive branches, which are nonspecific and specific, i.e. they target common and unique parts of non-self antigens, respectively (Table 1). They discriminate between self and non-self, and activate appropriate effectors. Although innate and adaptive immunity appear independent, appropriate interactions between them are indispensable for the normal function of the immune response. Indeed, "mistakes" that happen in either branch can affect the organism as a whole.

Innate immunity is the first defense against invading pathogens and performs the function of immune surveillance (Calandra et al., 2003; Janeway, 2001). Cells that constitute innate immunity are antigen-presenting cells (APC), including DC, macrophages and B cells. Innate immunity does not develop any antigen specificity because the discrimination between self and non-self is primarily based on pathogen associated molecular patterns (PAMP), which are common components of many microorganisms and are not found in humans, i.e. lipopolisacharides (LPS). Ligation of PAMP with toll-like receptors (TLR) on the surface of APC (Bachmann and Kopf, 2002; Takeda et al., 2003) leads to phagocytosis and APP. APP results in the presentation of processed antigens to T cells (Cresswell and Lanzavecchia, 2001) (Fig.1) and the establishment of adaptive immunity (Kelly et al., 2002; Smyth et al., 2001).

The function of adaptive immunity is the elimination of non-self antigens and the creation of immune memory. Constituents of adaptive immunity are B and T cells. Adaptive immunity is unique and created only after the encounter with a specific non-self antigen, which is presented to them by APC. Discrimination between self and non-self at the level of adaptive immunity is complex and involves the elimination or functional inactivation of self-reactive lymphocytes from the repertoire. Establishment of adaptive immunity is slow, but once it is established, it is memorized and able to respond faster to subsequent contacts with the same antigen. However, since innate immunity is faster, it keeps infections under control until the establishment of adaptive immunity (Smyth et al., 2003).

Very important role in APP, which connects innate and adaptive immunity, play the MHC determinants. Namely, processed antigenic peptides are presented to T cells only in the groove of MHC heterodimers.

## MAJOR HISTOCOMPATIBILLITY COMPLEX DETERMINANTS

Genes that encode MHC determinants, also known as human leukocyte antigens (HLA), are located on the short arm of chromosome six and are extraordinary polymorphic (anonymous, 1999). They are divided into two classes; MHC I and MHC II determinants that present intracellular and extracellular antigens, respectively. MHC I determinants are expressed on most nucleated cells, whereas MHC II determinants only on APC, mature B and activated T cells.

Three different MHC I determinants, namely HLA-A, HLA-B and HLA-C, are assembled and loaded with antigenic peptides, which are generated by protein degradation in the 26S proteasome, in the ER (Fruci et al., 2003; Saveanu et al., 2002). The complex between MHC I heterodimer and antigenic peptide passes through the trans-Golgi network to the cell surface of APC and activates CD8+ T cells.

In humans, there are three classical MHC II determinants termed HLA-DP, HLA-DQ and HLA–DR, and two non-classical determinants named HLA-DM and HLA–DO (Fig. 2). The former are cell surface heterodimers that present antigenic peptides, whereas the latter are cytoplasmic oligomers that are involved in loading of antigenic peptides. HLA-DP, HLA-DQ, HLA–DR and HLA-DM are composed of α and β chains, whereas HLA-DO form heterotetramers, composed of two α and two β chains (Fig. 2). Classical MHC II determinants are assembled in the ER in the complex with the invariant chain (Ii), which stabilizes them and prevents premature loading of antigenic peptides. This complex is transported to the late endosome that contains degraded antigens. Loading of antigenic peptides is a three-step process (Fig. 2). The first two steps, which take place in the late endosome, are the degradation and immediate replacement of Ii by the class II-associated Ii-chain peptide (CLIP). The third step happens in the MHC class II compartment (MIIC) and is the exchange of CLIP with antigenic peptides. This exchange is mediated by HLA-DM and is regulated by HLA-DO. The complex between MHC II determinants and antigenic peptides then travels to the surface of APC, where the antigen is presented to CD4+ T cells, which are the main regulators of the immune response.

Since they present antigens to CD4+ T cells, MHC II determinants are involved directly in the regulation of the immune response. Therefore, it is not surprising that the precise control of MHC II gene expression, which takes place at the level of transcription, is necessary for the equilibrated function of the immune system. Most of our knowledge about this control has been elucidated from studies of the severe combined immunodeficiency called MHC II deficiency or the type II bare lymphocyte syndrome (BLS) (Table 2).

## BARE LYMPHOCYTE SYNDROME

BLS is an autosomal recessive disease characterized by the lack of constitutive and inducible MHC II gene expression (Table 2) (reviewd in (Reith and Mach, 2001) and (Nekrep et al., 2003)). It is caused by mutations in factors that direct the transcription from MHC II promoters rather than in MHC II genes themselves. More than ten years ago, transcription of MHC II genes was known to require several unidentified proteins. The earliest studies performed to identify these proteins employed *in vitro* assays, which revealed many irrelevant DNA-protein interactions. However, only genetic studies of BLS finally resolved this puzzle. First, CIITA (Steimle et al., 1993) and regulatory factor X 5 (RFX5) (Steimle et al., 1995) were identified, which rescued the expression of MHC II determinants in complementation group (CG) A and C, respectively. Subsequently these findings led to the identification of regulatory factor X that contains ankyrin repeats (RFXANK/B) (Masternak et al., 1998; Nagarajan et al., 1999) and regulatory factor X associated protein (RFXAP) (Durand et al., 1997) that rescued the expression of MHC II determinants in CG B and D, respectively. These factors not only account for four CG of BLS, but also represent all gene-specific *trans*-acting factors that bind *cis*-acting elements for MHC II transcription.

## *cis*-ACTING ELEMENTS

MHC II genes are transcribed from compact promoters (Fig. 3A). They contain variable proximal promoter (PPS) and conserved upstream sequences (CUS) (reviewed in (Nekrep et al., 2003; Reith and Mach, 2001; Ting and Trowsdale, 2002)).

In HLA-DRA PPS are located from position – 52 to the transcription start site. From the 5' direction, they contain octamer binding site (OBS) and initiator (Inr) sequences, but lack a functional TATA box. OBS binds the octamer binding protein-1 (Oct-1), which recruit the B cell octamer binding protein 1/Octamer binding factor 1/Oct coactivator from B cells (Bob1/OBF-1/OCAB) (Matthias, 1998), whereas the initiator, which represents the transcription start site, binds RNA polymerase II associated transcription factor I (TF$_{II}$I) and TATA binding protein (TBP) associated factor of 250 kDa (TAF$_{II}$250).

CUS are located from positions –139 to –67. From the 5' direction, they contain S, X and Y boxes. X box can be further divided into X1 and X2 boxes. Tight spatial constraints are preserved between these boxes (Jabrane-Ferrat et al., 1996), which bind different *trans*-acting factors. S and X1 boxes bind RFX (Table 2) (Jabrane-Ferrat et al., 1996; Ting and Trowsdale, 2002), the X2 box binds X2 binding protein (X2BP), which can be c-AMP responsive element binding protein (CREBP) or activator protein 1 (AP-1) (Moreno et al., 1997; Setterblad et al., 1997) and the Y box binds nuclear factor Y (NF-Y) (Table 2) (Maity and de Crombrugghe, 1998; Mantovani, 1999).

However, PPS and CUS are not sufficient for HLA-DRA gene expression. A distal locus control region (LCR), which is located approximately 2.4 kb upstream of the CUS, is needed for efficient transcription from the HLA-DRA promoter in the organism (Masternak et al., 2003) (Fig. 3). LCR is a mirror image of CUS. It contains S', X1', X2' and Y' boxes, but in the opposite orientation from the promoter. Indeed, the LCR binds the same *trans*-acting factors as the CUS.

## *trans*-ACTING FACTORS

Several *trans*-acting factors are required for MHC II transcription (Fig. 4). Among them, RFX and NF-Y create the platform for other co-activators to access MHC II promoters.

NF-Y is composed of three subunits (Table 2) (reviewd in (Matuoka and Yu Chen, 1999)). NF-YA contains 374 residues. At the N-terminus of NF-YA, there is glutamine-rich region, which is followed by a stretch of prolines, serines and threonines. Its subunit interaction and DNA binding domains (DBD) are at the C-terminus.

NF-YB contains 207 residues. They form a histone fold and TATA-binding protein (TBP) binding domains. The histone fold is similar to that of histones 2A and 2B (Mantovani, 1999), which is responsible for their dimerization. This domain has the same function in the assembly of NF-Y.

NF-YC contains 335 residues. As with NF-YB, it contains a histone fold and TBP binding domains, and in addition, a glutamine-rich region.

RFX is also composed of three subunits (Table 2) (reviewd in (Reith and Mach, 2001)). RFXANK/B contains 260 residues that form an acidic domain and four ankyrin repeats, which gave it its name.

RFX5 contains 616 residues, which form DBD and PEST domains. It is the 5th member of the RFX family that binds S and X boxes of MHC II promoters, which gave it its name.

RFXAP contains 272 residues. This protein has acidic, basic and glutamine-rich domains. Because it interacts directly with RFX5, it was named the RFX associated protein.

Transcription from MHC II promoters starts after productive interactions between *cis*-acting elements and *trans*-acting factors, which lead to the formation of the MHC II enhanceosome and transcriptosome.

## MHC II ENHANCEOSOME AND TRANSCRIPTOSOME

The MHC II enhanceosome is formed on CUS and LCR. Its assembly requires specific protein-protein and DNA-protein interactions. The formation of the enhanceosome starts off DNA (Fig. 5) where RFX and NF-Y bind loosely. Both RFX and NF-Y are formed in two steps. RFXANK/B forms a complex with RFXAP, which binds the RFX5 oligomer (Jabrane-Ferrat et al., 2002). NF-YB and NF-YC first form heterodimer via their histone folds and then bind NF-YA. RFX and NF-Y are linked via interactions between RFXANK/B, NF-YA and NF-YC (Jabrane-Ferrat et al., 2002; Nekrep et al., 2003; Ting and Trowsdale, 2002). The binding of the enhanceosome to promoters occurs via DBD of RFX5, NF-YB and NF-YC. Each individual contact between the enhanceosome and DNA takes place on the same side of the double helix. As mentioned earlier, preserved spatial constraints are present in CUS and LCR. Completely invariant spacing between S and X boxes can be explained by the RFX5 oligomers, which bind each others next to theirs DBD. These higher ordered RFX complexes can occupy S and X boxes only if the exact spacing between them is preserved. Between X and Y boxes are tolerated full helical turns. This finding can be explained because their binding occurs far from DBD.

NF-Y not only interacts with RFX, but also with different histone acetyltransferases (HAT) that make chromatin accessible for other co-activators (Fontes et al., 1999; Harton et al., 2001). Moreover, NF-YB and NF-YC bind TBP *in vitro*. This interaction helps to attract general transcription factors (GTF) to promoters, which lack a TATA box. The X2 box is bound by CREBP or AP-1 (Moreno et al., 1995; Setterblad et al., 1997).

The very last step is the recruitment of CIITA to the MHC II enhanceosome, which converts it into the MHC II transcriptosome (Fig. 5). Once the transcriptosome is formed MHC II genes can be transcribed.

# CLASS II TRANSACTIVATOR

CIITA is the master regulator of MHC II gene expression. Except for CIITA, all *trans*-acting factors that are needed for the transcription of MHC II genes are expressed ubiquitously. Thus, the synthesis of MHC II determinants correlates directly with the presence of CIITA, which is expressed constitutively only in DC and mature B cells and is inducible in most other cell types. After CIITA is recruited to the MHC II enhanceosome, it recruits the general transcriptional machinery.

CIITA contains 1130 residues (Fig. 7). It can be divided into several domains, which are the N- terminal activation (AD), proline/serine/threonine rich (PST) and GTP binding domains (GBD), as well as the C-terminal leucine rich region (LRR). Additionally, CIITA contains three nuclear localization signals (NLS) and two putative nuclear export sequences (NES) (Fig. 7).

The optimal AD of CIITA spans the first 322 residues. Its amino terminal part is rich in aspartic and glutamic acids and resembles the classical acidic AD, similar to that of VP16. Indeed, it can be replaced by AD from VP16 (herpes simplex) and E1a (adenovirus) (Riley and Boss, 1993). Between residues 145 and 322 is the PST domain. As its name implies, it contains several prolines, serines and threonines that are targets for post-translational modifications. A consensus PEST sequence is located from positions 283 to 308, but it does not represent a degradation motif or degron (Schnappauf et al., 2003). Rather, degrons are located within the first 99 residues and from positions 230 to 260 of CIITA, respectively (Schnappauf et al., 2003).

A putative GBD resides downstream of AD, from positions 420 to 561. In contrast with the classical GBD, which contain four GTP-interacting sequences, CIITA contains only three (Harton et al., 1999). These are Walker type A motif (G1), magnesium coordination site (G3) and a site that confers specificity for guanosine (G4). Indeed, although CIITA binds GTP, it does not hydrolyse it *in* *vitro* (Harton et al., 1999). This finding suggests that CIITA might be a constitutively active GTP-binding protein. As mutations of either motif reduce the activity of CIITA, the sole function of GTP binding to CIITA may be to alter its conformation.

Five or six consensus LRR are located from positions 988 to 1097 (Harton et al., 2002). They bind a 33 kDa protein of unknown function (Hake et al., 2000). Certain mutations positioned in α helices of LRR, but not in β sheets decrease the activity of CIITA. An additional LRR flanking the GBD might be involved in the aggregation of CIITA (Linhoff et al., 2001).

In CIITA, three NLS and two NES have been described. A bipartite NLS is located from positions 141 to 159, and two additional NLS were mapped from positions 405 to 414 next to G1 and 940 to 963 in the LRR (Cressman et al., 1999; Cressman et al., 2001; Nekrep et al., 2002; Spilianakis et al., 2000). NES have been poorly characterized. The first is located in the first 114 residues, the second from positions 408 to 550 (Kretsovali et al., 2001). However, none of these NLS and NES have been examined directly.

## REGULATION OF CIITA GENE EXPRESSION

Since the transcription of MHC II genes is dependent on the presence of CIITA, it is not surprising that its expression is highly regulated. In general, there are two types of regulation of gene expression: genetic and epigenetic. CIITA employs both. Whereas specific activators dictate active transcription, epigenetic silencing occurs via chromatin condensation.

Transcription of CIITA can be initiated from up to three different promoters: PI, PIII and PIV (Fig. 6) (Muhlethaler-Mottet et al., 1997). In DC and mature B cells, constitutive expression of CIITA is initiated from PI and PIII, respectively (Muhlethaler-Mottet et al., 1997). IFN-γ inducible expression of CIITA is mediated by PI and PIV in bone marrow derived APC, DC and somatic cells, respectively (Muhlethaler-Mottet et al., 1997; Waldburger et al., 2001). All promoters have a unique exon 1, which is spliced into the common exon 2 (Muhlethaler-Mottet et al., 1997). Translation of CIITA transcripts from PI and PIII starts from the first methionine in exon 1, whereas from PIV it begins from the first methionine in exon 2. This brings unique sequences to the N-terminus of CIITA (Muhlethaler-Mottet et al., 1997) and results in three isoforms (IF) of CIITA: IF I, IF III and IF IV. IF I and IF III contain additional 94 and 17 residues, respectively. The additional sequence in IF I bears homology with the caspase recruitment domain (CARD) (Nickerson et al., 2001) and most likely represents a new protein-protein interaction domain. It also has the highest transcriptional activity (Nickerson et al., 2001). Importantly, a single mutation of a conserved leucine in CARD abrogates the activity of CIITA (Nickerson et al., 2001).

Epigenetic regulation of CIITA gene expression is utilized for embryonic survival. Notably, immune responses against paternal antigens in the placenta cause fetal death. In these tissues, promoter hypermethylation and histone deacetylation are proposed mechanisms for silencing the transcription of CIITA (Holtz et al., 2003; Morris et al., 2000; van den Elsen et al., 2000). Derepression of CIITA transcription in trophoblasts after treatment with methylation inhibitors and trichostatin A with IFN-γ supports this mechanism.

In immature DC, the expression of CIITA is observed from PI, but it does not lead to APP. Phagocytosis of an antigen leads to the maturation of DC, which is accompanied by the cessation of further phagocytosis and repression of *de novo* CIITA and MHC II transcription and up-regulation of APP (Cella et al., 1997; Landmann et al., 2001; Pierre et al., 1997; Turley et al., 2000). These events might be mediated by TLR. Because only the specific population of encountered antigens are processed and presented to T cells, such regulation could represent the best way of establishing adaptive immunity. Because the expression of CIITA is shut down after the clearance of a foreign antigen, it also lowers the chance of DC randomly presenting self-antigens.

In B cells, the constitutive expression of CIITA from PIII is accompanied by the presence of high levels of MHC II determinants on the cell surface. However, after B cells differentiate into highly specialized antibody producing plasma cells, the expression of CIITA and MHC II determinants is lost (Chen et al., 2002). B lymphocyte inhibitory maturation protein 1 (BLIMP1) begins the process of this silencing (Piskurich et al., 2000).

CIITA gene expression can be induced with IFN-γ from PI and PIV. PIV contains gamma-activated sites (GAS), which bind the signal transducer and activator of transcription 1 (STAT1) and an interferon regulatory factor-1 binding site, which binds interferon regulatory factor-1 (IRF-1) (Muhlethaler-Mottet et al., 1998). Since PI is also responsive to IFN-γ, it could contain the same elements. After stimulation with IFN-γ, activated, phosphorylated STAT 1 translocates into the nucleus and binds GAS. Activated STAT 1 also induces the expression and accumulation of IRF-1. Binding of STAT 1 to GAS itself causes a weak acetylation of histones 3 and 4 in PIV (Morris et al., 2002). However, the hyperacetylation of histones, which makes transcription more efficient, happens only after the accumulation and binding of IRF-1 to its site.

Thus, the regulation of transcription of the CIITA gene represents the first level of control of CIITA function. The second level consists of post-translational modifications of CIITA.

## POST-TRANSLATIONAL MODIFICATIONS OF CIITA

CIITA is post-translationaly modified by acetylation, phosphorylation and ubiquitylation (Fig. 8). Effects of these modifications are complex and far from being clearly understood, but a general picture can be drawn from several studies. The fate of CIITA in the cell is not random, but each step, starting after its translation and ending with its possible degradation, is regulated precisely by the following modifications.

Phosphorylation is the first post-translational modification of CIITA. Two studies have shown that this modification increases the activity of CIITA. The first study mapped phosphorylation sites into the PST region from positions 253 to 321 (Tosi et al., 2002). The second study showed that protein kinase A (PKA) phosphorylates CIITA on one or more serines in the region between PST and GBD (Sisk et al., 2003). The phosphorylation of CIITA leads to its accumulation, oligomerization and nuclear translocation. Most likely, these latter events happen because the phosphorylation changes the conformation of CIITA and exposes its NLS.

After it translocates into the nucleus, CIITA is acetylated on lysines 141 and 144, by CBP/p300 and also by pCAF, which are located in the nucleus (Spilianakis et al., 2000). Acetylation of these residues keeps CIITA in the nucleus and increases the stability of the MHC II transcriposome. Acetylation might also facilitate the subsequent ubiquitylation of CIITA. Indeed, recent data suggest that histone deacetylases (HDAC) are involved in the ubiquitylation of CIITA (Greer et al., 2003).

Ubiquitylation, which is a covalent modification of lysines, plays an important role in transcription. Monoubiquitylated transcription factors tend to be more potent activators, whereas polyubiquitylated proteins are destined for degradation by the 26S proteasome. The role of ubiquitylation of CIITA has been addressed in two studies, but clear conclusions on the function of ubiquitin cannot be reached. In the first study, monoubiquitin fused to CIITA prevented the degradation of the modified CIITA protein, but did not affect its transcriptional activity (Schnappauf et al., 2003). In the second study, CIITA co-expressed with ubiquitin had a higher activity than CIITA alone, which was even more pronounced if ubiquitin could not form oligomers (Greer et al., 2003). This effect of ubiquitin was on the stabilization of the MHC II transcriptosome rather than on transcription itself. Interestingly, the same study showed that HDAC prevent the ubiquitylation of CIITA.

Phosphorylation also completes the post-translational modifications of CIITA. Interestingly, PKA inactivates CIITA with phosphorylation of serines 834 and 1050. (Li et al., 2001). Does this modification influence the stability of the enhanceosome? Does it lead to the degradation of CIITA, or does it expose the NES and enables the export of CIITA from the nucleus?

## CIITA IS SHUTTLING PROTEIN

Transcription factors and co-factors need to be transported into the nucleus to perform their function. Nucleo-cytoplasmic shuttling usually requires two types of functionally unique signals, NLS and NES. CIITA contains both, but none of them have been examined adequately (Fig. 7).

At steady state, CIITA is distributed equally between the nucleus and the cytoplasm. Mutant CIITA proteins, which do not follow this pattern, led to the identification of NLS. As mentioned earlier, two NLS are located at the N-terminus and one NLS in the C-terminus of CIITA (Fig. 7) (Cressman et al., 1999), (Cressman et al., 2001), (Spilianakis et al., 2000). However, none of them has been examined directly, i.e. they have not been demonstrated to shuttle a heterologous protein. Moreover, binding to importins has not been investigated.

Two regions in the N-terminus of CIITA have been proposed to function as NES (Fig. 7). The only support comes from interactions between these regions with the chromosomal region maintenance-1 (Crm-1) protein (Kretsovali et al., 2001). Indeed, treatment with leptomycin B (LMB), which blocks Crm-1 dependent nuclear export (Kudo et al., 1998), leads to the nuclear accumulation of CIITA (Cressman et al., 2001; Kretsovali et al., 2001). However, no consensus NES can be found within these sequences. In addition none of the five putative NES, three of which correspond perfectly to the consensus sequence [Lx(2,3)Lx(2,3)-LxL], functioned in a direct export nor bound Crm-1 (Drozina, Kohoutek, Peterlin, unpublished observation).

All these results imply the presence of other transport mechanism/s. Indeed, several additional regions of CIITA are involved in its nuclear localization, including GBD (Harton et al., 1999; Raval et al., 2003) and LRR (Hake et al., 2000; Harton et al., 2002; Towey and Kelly, 2002). Moreover, some mutations in the LRR disrupt interaction between p33 and CIITA, which interferes with its nuclear localization (Hake et al., 2000). Furthermore, it was suggested that NES from CIITA resemble that of snurportin-1, which is discontinuous (Paraskeva et al., 1999). In this case, only after a conformational change can Crm-1 bind and transport snurportin-1 from the nucleus to the cytoplasm (Paraskeva et al., 1999).

## CIITA IS TRANSCRIPTIONAL INTEGRATOR

Once CIITA is modified appropriately and present in the nucleus, it can perform its function. It binds to the MHC II enhanceosome, attracts several transcription factors and co-factors and integrates initiation and elongation of transcription as well as chromatin remodeling into a process that finally results in the expression of MHC II genes. Moreover, CIITA might also be involved in the dissociation of the MHC II enhanceosome after the termination of transcription.

CIITA recruits RNA polymerase II (RNAPII) to MHC II promoters and increases rates of initiation and elongation of transcription. For the former function, interactions with TAF (Fontes et al., 1997), TFIIB (Mahanta et al., 1997) and Bob1/OBF-1/OCAB (Fontes et al., 1996) might be necessary. For the latter, the interaction between CIITA and the positive transcription elongation factor b (P-TEFb), which phosphorylates the CTD of RNAPII, is needed (Kanazawa et al., 2000).

Not surprisingly, CIITA also recruits HAT and chromatin remodeling machinery to the HLA-DRA promoter. The AD of CIITA binds CBP/p300 (Fontes et al., 1999), pCAF (Spilianakis et al., 2000) and Brahma-related gene 1 (BRG-1) (Mudhasani and Fontes, 2002). In addition, it has been suggested that CIITA posseses intrinsic HAT activity (Harton et al., 2001; Raval et al., 2001). This activity of CIITA has been mapped into AD and bears homology with other HAT domains, e.g. in CBP/p300 (Harton et al., 2001). Interestingly, CIITA that lacks its AD is still able to mediate the acetylation of histone 4, but not of histone 3 (Beresford and Boss, 2001). Thus, CIITA could contain another region with direct or indirect HAT activity.

After the termination of transcription, deacetylation of histones, which is mediated by HDAC, is involved in chromatin condensation. Binding of CIITA to HDAC1 and HDAC2 decreases the activity of CIITA, most probably because histone deacetylation disrupts the MHC II enhanceosome (Zika et al., 2003). Thus, CIITA might play an active role in the regulation of its own function on MHC II promoters.

## CONCLUSIONS AND FUTURE DIRECTIONS

In the last few years, much research resulted in greater understanding of the function and regulation of CIITA. Since CIITA plays a major role in the regulation of the immune response via MHC II determinants, this is not surprising. Moreover, if one understood the "rules" of this regulation, one could use this knowledge for the manipulation of the immune system. However, many questions remain.

Why is CIITA transcribed from three distinct promoters? Are different IF of CIITA modified differently, which influences their activity and regulation? Are post-translational modifications independent from each other or are they carefully orchestrated or sequential? The phosphorylation of CIITA increases or decreases the activity of CIITA, depending on phosphorylated residues. What signals regulate this phosphorylation that results in activation and inactivation of CIITA, respectively, and how? Why does the acetylated CIITA protein accumulate in the nucleus? What is the role of its ubiquitylation? Where is CIITA ubiquitylated? These and probably many more questions regarding the post-translational modifications of CIITA still need to be answered.

The shuttling of CIITA is a big puzzle. Since CIITA has not been shown to bind importins and since no NES can be found, how does CIITA shuttle? And why does it shuttle at all?

A lot is known about the *cis*-acting elements in the MHC II promoters, but the role of LCR needs more attention. What is the mode of action of these LCR? Do they also bring P-TEFb to RNAPII? How many LCR are there in the MHC cluster?

Once these questions have been answered, the manipulation of the immune system could become possible and eventually applicable in clinical medicine. Cancer and inadequate vaccination represent weak immune responses. Introduction of the constitutively active CIITA proteins could turn cancer cells into APC. They would then present their own antigenic peptides to CD4+ T cells to activate and direct immune response against transformed cells. Vaccination could also be more effective with the addition of CIITA. Autoimmunity is caused by the inappropriate responses to self-antigens. Dominant negative CIITA proteins could attenuate immune responses as well as the need for other immunosuppressive therapies. Thus, CIITA could become a new tool for immunotherapy in humans.

**Figure 1.**

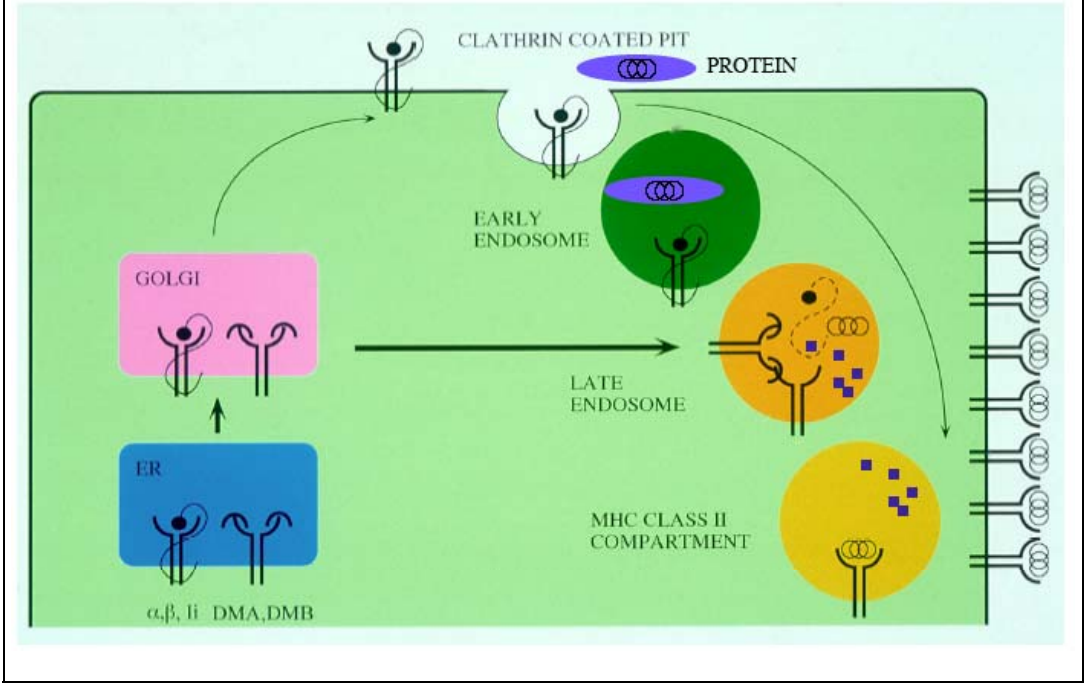

**Figure 2.**

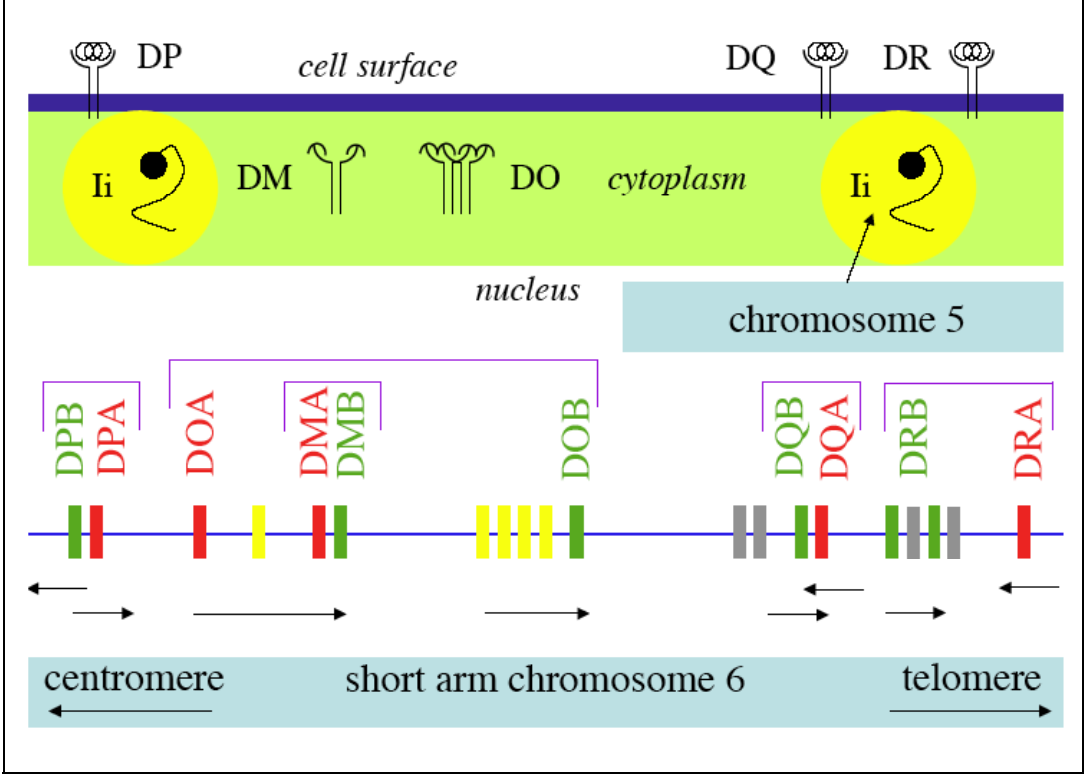

# Figure 3.

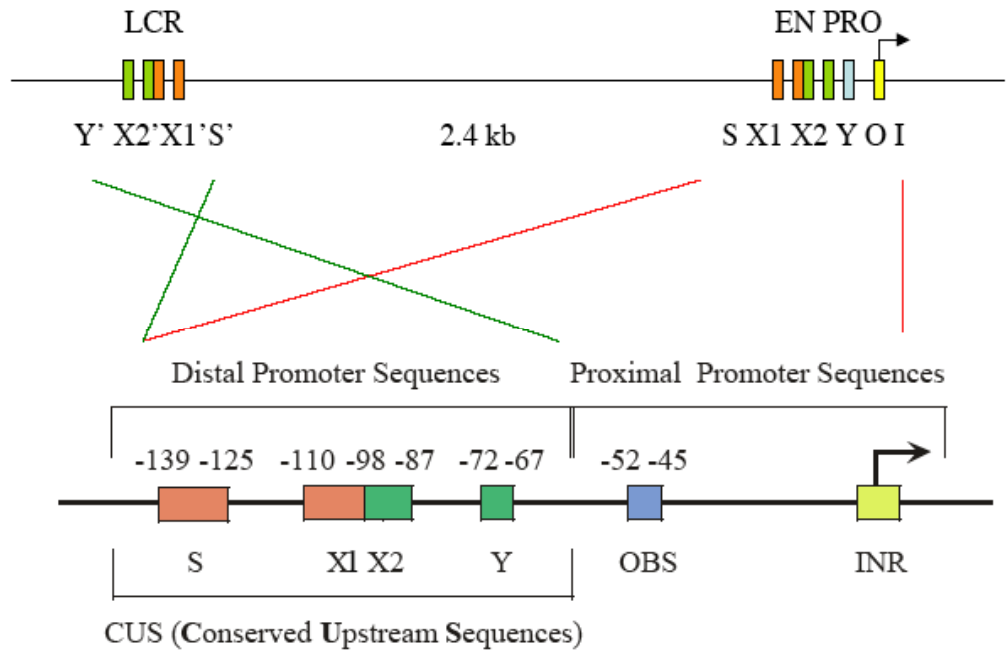

# Figure 4.

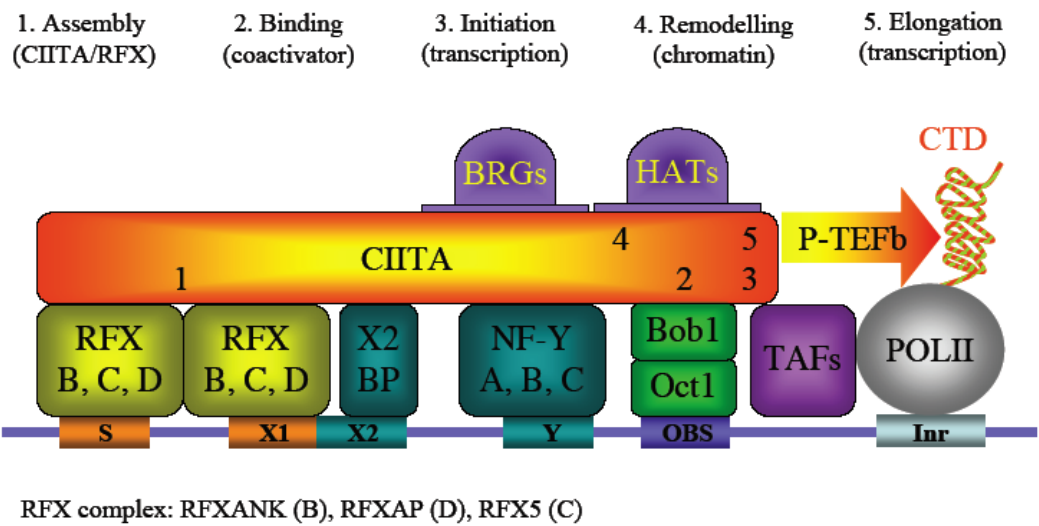

# Figure 5.

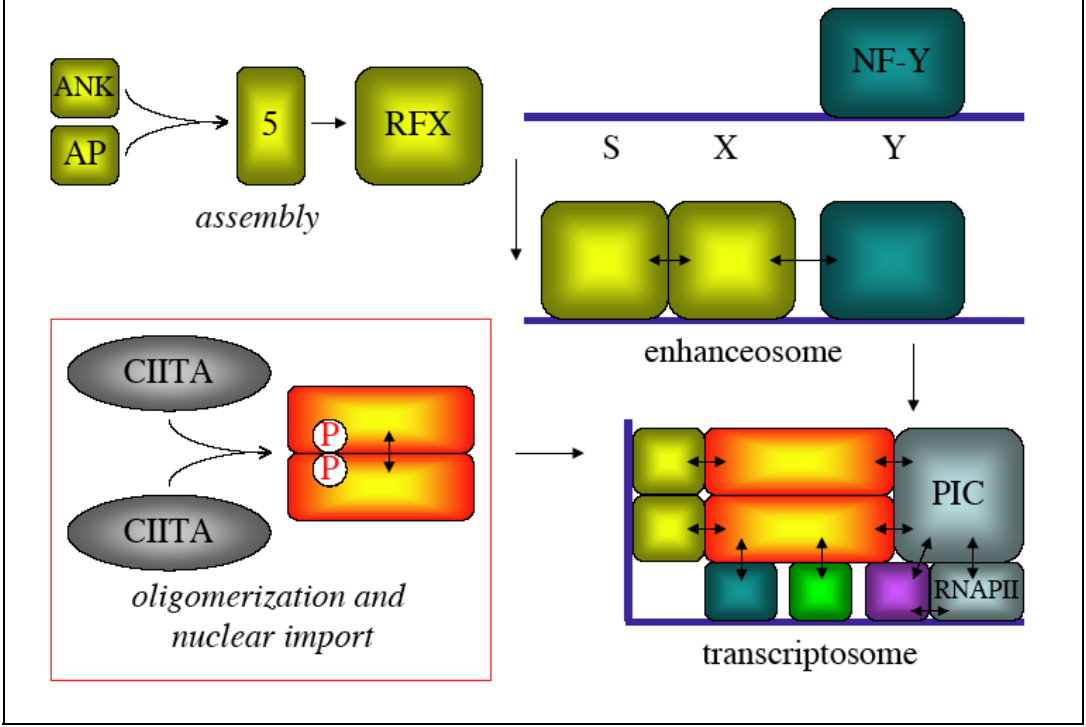

# Figure 6.

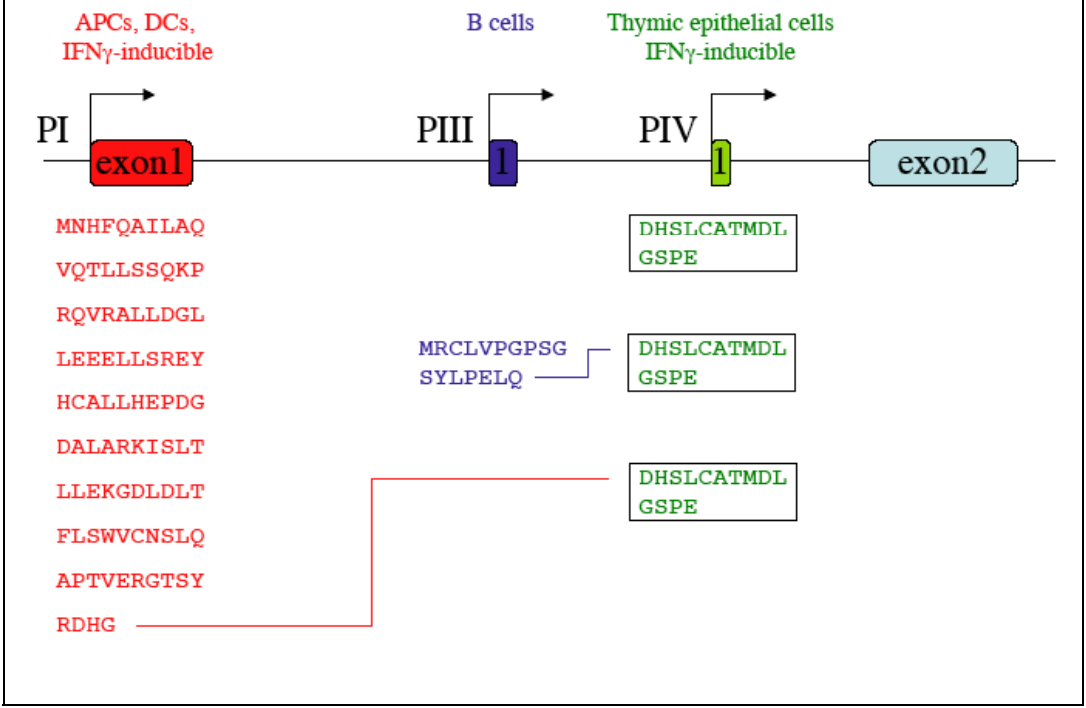

# Figure 7.

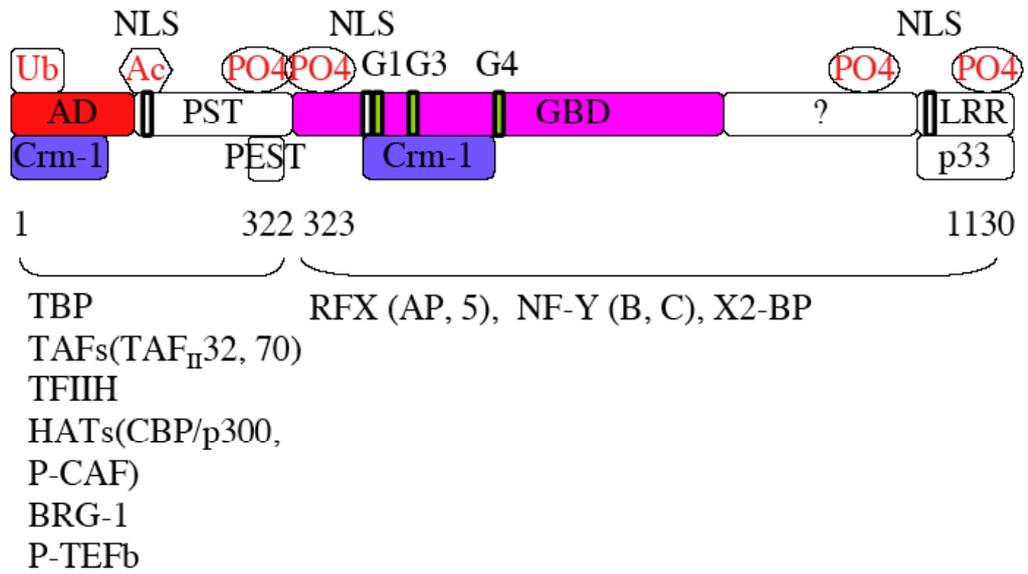

# Figure 8.

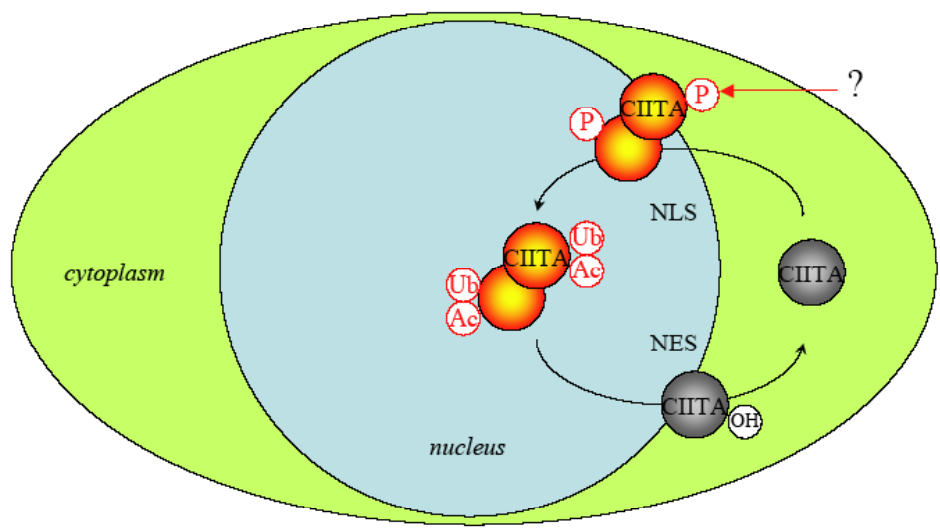

**Table 1.**

|  | *INNATE* | *ADAPTIVE* |
|---|---|---|
| *Tolerance to self* | Yes | Yes |
| *Specificity* | Low | High |
| *Diversity* | Low | High |
| *Memory* | None | Yes |
| *Cells* | APC | B and T cells |

Abbrevation: APC; antigen presenting cells.

**Table 2.**

| *Complementation group* | *A* | *B* | *C* | *D* |
|---|---|---|---|---|
| *MHC II expression* | Absent | Absent | Absent | Absent |
| *Enhanceosome* | Formed | Not formed | Not formed | Not formed |
| *Mutated gene* | *MHC2TA* | *RFXANK/B* | *RFX5* | *RFXAP* |
| *Mutated protein* | CIITA | RFXANK/B | RFX5 | RFXAP |

Abbrevations: *MHC2TA;* major histocompatibility class 2 trans-activator, RFXANK/B; regulatory factor X ANK/B, RFX5; regulatory factor X 5, RFXAP; regulatory factor X AP, CIITA; class II transactivator.

FIGURE LEGENDS

**Figure 1.** *Pathway of antigen processing and presentation (APP)*
HLA-DP, HLA-DQ, HLA-DR, which are in complex with Ii, and HLA-DM heterodimers are assembled in the endoplasmic reticulum (ER, depicted in blue). They travel to the trans Golgi network (depicted in pink) and either fuse with the late endosome (depicted in orange) or continue to the cell surface. Degradation of phagocytozed antigenic peptides (depicted in violet) starts in an early endosome (depicted in green). In the late endosome, Ii is removed from the complex with HLA-DP, HLA-DQ, HLA-DR and replaced with antigenic peptides. Loaded with antigens, MHC II molecules travel from MHC class II compartment (MIIC, depicted in yellow) to the cell surface, where they activate T cells.

**Figure 2.** *The organization and direction of MHC II genes, as well as sub-cellular localization of their proteins*
Three heterodimers are transcribed from promoters that point in opposite directions (HLA-DPA and HLA-DPB, HLA-DQA and HLA-DQB, HLA-DRA and HLA-DRB), HLA-DO heterodimer and HLA-DM heterotetramer are transcribed from promoters that point towards the telomere. Genes for α and β chains are shown as red and green bars, respectively. Pseudogenes and MHC II unrelated genes are shown as grey and yellow bars, respectively. Black arrows point in the direction of transcription.

HLA-DP, HLA-DQ and HLA-DR heterodimers are found on the cell surface in complex with antigenic peptides. HLA-DM heterodimers and HLA-DO heterotetramers are cytoplasmic molecules, as well as invariant chain (Ii), all of which are also found in MHC class II compartments (MIIC, depicted as yellow circles). Genes coding for HLA-DP, HLA-DQ, HLA-DR, HLA-DM and HLA-DO are located on the short arm of chromosome 6, but gene coding for Ii is located on the chromosome 5.

**Figure 3**. *cis-acting elements in the DRA promoter*
Locus control region (LCR), conserved upstream sequences (CUS), and proximal promoter sequences (PPS) are needed for the transcription from the DRA promoter. LCR is located 2.4 kb upstream from CUS. LCR and CUS contain Y', X2', X1', S' and S, X1, X2 and Y boxes, respectively. PPS contain the octamer binding site (OBS) and the initiator (INR).
The spacing between CUS and PPS are presented in the lower pannel.

**Figure 4.** *cis-acting elements and trans-acting factors on MHC II promoters*
Conserved upstream sequences (CUS) contain S, X and Y boxes. X box is divided into X1 and X2 boxes. In DRA gene, the octamer binding site (OBS) and the initiator (INR) form the proximal promoter sequences (PPS). RFX, which contains RFXANK, RFXAP and RFX5, binds S and X1 boxes. X2 binding protein (X2BP) binds X2 box. NF-Y, which contains NF-YA, NF-YB and NF-YC, binds Y box. OBS binds octamer binding protein-1 (Oct-1), which recruit the B cell octamer binding protein 1/Octamer binding factor 1/Oct coactivator from B cells (Bob1/OBF-1/OCAB). The initiator binds general transcription factors and positions RNA polymerase II (RNAPII) on the promoter. CIITA is recruited to the MHC II enhanceosome, which consists of RFX, X2BP and NF-Y on CUS. It binds histone acetyl-transferases (HAT, CBP/p300, pCAF) and Brahma related gene (BRG) proteins that remodel the chromatin, Bob1/OBF-1/OCAB, GTF and TAF that recruit RNAPII and initiate transcription, and positive transcription elongation factor b (P-TEFb), that phosphorylates the C terminal domain (CTD) of RNAPII and facilitates the elongation of transcription. These steps in the transcriptional process are given above the drawing.

**Figure 5.** *Assembly of the MHC II enhanceosome and transcriptosome*
RFXANK/B forms complex with RFXAP, which recruits the RFX5 oligomer. RFX binds S and X1 boxes, whereas NF-Y binds Y box. They form the MHC II enhanceosome. The phosphorylation of CIITA leads to its oligomerization and nuclear import. It binds to the MHC II enhanceosome and forms the MHC II transcriptosome, which forms the preinitiation complex (PIC) and recruits as well as modifies RNA polymerase II (RNAPII).

**Figure 6.** *CIITA promoters, their composition and specificity*
CIITA is transcribed from three different promoters: PI, PIII and PIV with unique exons 1 (depicted as red, purple, and green boxes for PI, PIII, and PIV, respectively) that are spliced into common exon 2 (depicted as a blue box). The additional sequences in CIITA, transcribed from PI is given in red letters and from PIII in purple letters, while a common sequence from all promoters is given in green letters. Cell types and stimuli are given above the scheme. Abbreviations: IFN-γ, interferon gamma; APC, antigen-presenting cells; DC, dendritic cells.

**Figure 7.** *The scheme of CIITA, its post-translational modifications and partner proteins*
CIITA is divided into several domains, which are activation (AD), proline/serine/threonine rich (PST), GTP binding (GBD) domains and leucine-rich repeats (LRR). Additionally, two nuclear localization signals (NLS) are located in the N-terminus and one in the C-terminus of the protein. Within PST domain there is a consensus proline/glutamine/serine/threonine rich (PEST) sequence. GBD contains three GTP interacting sequences, designated by G1, G3 and G4. Positions of post-translational modifications of CIITA, which are phosphorylation, acetylation and ubiquitylation, are given above the drawing in red letters as PO4, Ac and Ub, respectively. Chromosomal region maintenence-1 protein (Crm-1, depicted as a purple box) interacts with AD and GBD, and p33 (depicted as a white box) binds LRR. Transcriptional co-activators (TBP, TAFs, HATs, BRG-1, P-TEFb) as well as constituents of the MHC II enhanceosome (RFXAP, RFX5, NF-YB, NF-YC and X2BP), which interact with CIITA, are listed under the drawing.

**Figure 8.** *Post-translational modifications of CIITA*
After an unknown signal (depicted as the red arrow), CIITA is phosphorylated, which causes its oligomerization and nuclear translocation (black arrow directed from the cytoplasm into the nucleus). Acetylation and ubiquitylation increase transcriptional activity of CIITA. Termination of transcription is accompanied by additional modifications of CIITA, possibly phosphorylation that might result in the translocation of CIITA into the cytoplasm (black arrow directed from the nucleus into the cytoplasm) or in its degradation. CIITA monomers and oligomers are depicted as gray and orange circles. Phosphorylation of CIITA is represented as a red P attached to CIITA. Acetylation and ubiquitylation of CIITA are depicted as red Ac and Ub, respectively. NLS and NES stand for nuclear localization signal and nuclear export sequence, respectively.

Table 1.   Properties of innate and adaptive immunity
Table 2.   Properties of bare lymphocyte syndrome